\documentclass[11pt,amsmath,floats,floatfix,nofootinbib,secnumarabic,superscriptaddress]{revtex4}
\usepackage{graphicx} 
\usepackage{bm} 
\usepackage{epsfig}
\usepackage{pstricks}
\usepackage{setspace}
\usepackage{bm}
\usepackage{color}
\usepackage{colordvi}
\newcommand{\beq}{\begin{equation}}
\newcommand{\eeq}{\end{equation}}
\newcommand{\bea}{\begin{eqnarray}}
\newcommand{\eea}{\end{eqnarray}}
\newcommand{\bit}{\begin{itemize}}
\newcommand{\eit}{\end{itemize}}

\newcommand{\nn}{\nonumber}

\newcommand{\beal}{\begin{align}}
\newcommand{\eal}{\end{align}}
\newcommand{\bes}{\begin{split}}
\newcommand{\es}{\end{split}}
\newcommand{\calo}{{\cal O}}

\begin{document}

\title{Spin induced multipole moments for the gravitational wave amplitude from binary inspirals to 2.5 Post-Newtonian order}
\author{Rafael A. Porto}
\email{rporto@ias.edu}
\affiliation{Institute for Advanced Study, Princeton, NJ 08540, USA}
\affiliation{Department of Physics \& ISCAP, Columbia University, New York, NY 10027, USA}
\author{Andreas Ross}
\email{andreasr@andrew.cmu.edu}
\affiliation{Carnegie Mellon University Dept. of Physics, Pittsburgh PA 15213, USA}
\author{Ira Z. Rothstein}
\email{izr@andrew.cmu.edu}
\affiliation{Carnegie Mellon University Dept. of Physics, Pittsburgh PA 15213, USA}
\affiliation{California Institute of Technology, Pasadena CA, 91125, USA}

\begin{abstract}
Using the NRGR effective field theory formalism we calculate the remaining source multipole moments necessary to obtain the spin contributions to the gravitational wave amplitude to 2.5 Post-Newtonian (PN) order. We also reproduce the tail contribution to the waveform linear in spin at 2.5PN arising from the nonlinear interaction between the current quadrupole and the mass monopole.
\end{abstract}
\maketitle
\section{Introduction}

Direct detection of gravitational waves (GWs) is an international experimental effort which includes the earth-based LIGO/Virgo detectors \cite{LIGOVIRGO} and the future space-based mission eLISA \cite{eLISA}. Among the most promising sources for these detectors are coalescing compact binary systems, whose dynamics can be studied using Post-Newtonian  methods for the inspiral phase and numerical simulations for the merger.

The search for GWs from compact binaries is performed by the method of matched filtering, which relies on having accurate theoretical templates for the expected GW signal.  Matched filtering for the periodic GW signals emitted by binary systems is most sensitive to the phase. That is the reason for the common use of restricted waveforms, which include only the leading PN effects in the amplitude but higher PN orders in the phase. However, a precise knowledge of the GW amplitude has several benefits: It increases the mass range through the inclusions of higher harmonics \cite{massrange}, improves parameter estimation \cite{parameterestimation},  yields a higher angular resolution and a better the measurement of the luminosity distance \cite{angularresolution}, and is necessary for the comparison of numerical and analytic results \cite{numericscomparison}.

Measurements of black hole spins indicate that black holes in binary systems may frequently be close to maximally rotating \cite{spinBHs}. Thus it is timely and necessary to develop more accurate GW templates that include spin effects to higher orders in both the phase and the amplitude. For nonrotating compact objects, the GW amplitude from binary inspirals is known to 3PN \cite{Blanchet:2008je}, but spin effects have only been fully included at 1.5PN order and partially at 2PN \cite{ampspin}. \\

In this paper we compute the spin effects in the GW amplitude to 2.5PN. We use the NRGR effective field theory (EFT) formalism developed in \cite{nrgr, nrgrs, nrgrproc, rad1, anditt, dis1, dis2} (see \cite{LH, nrgr1} for reviews), with which the conservative part of the dynamics for rapidly rotating compact objects has been completed to 3PN order in \cite{andi1, foffa, eih,comment,nrgrss,nrgrs2,nrgrso,perrodin,Levi:2010zu}. (The next-to-leading order (NLO) spin-orbit and spin-spin Hamiltonians have been also computed in \cite{buo1,Schafer3pn,Schafer3pn2} using more traditional PN methods, and full agreement was reported in \cite{Hergt}. The conservative spin(1)-spin(2) potential was recently computed at NNLO using both the EFT \cite{Levi:2011eq} and the ADM formalism \cite{Hartung:2011ea}.)

Blanchet, Buonanno and Faye computed in \cite{buo2} the source multipole moments linear in spin for the spin effects in the energy flux and the GW phase to 2.5PN in a different framework and with different conventions from ours. In \cite{srad} we calculated the source multipole moments needed to obtain the spin contributions to the power emitted in GWs to 3PN order. These results include the corrections linear in spin, previously obtained in \cite{buo2}, and  the hitherto uncalculated spin(1)-spin(1) and spin(1)-spin(2) contributions. The latter provided the last ingredients required to determine the GW phase evolution to 3PN order, which will be presented in \cite{powerloss} where we will include a detailed comparison to the results of \cite{buo2}. 

In the present paper we extend our work in \cite{srad} and calculate the missing ingredients to obtain the spin effects in the gravitational waveform to 2.5PN order, which were not computed in \cite{srad} since they enter the GW phase beyond 3PN.
The source multipole moments calculated here are linear in spin, and consist of the LO current and mass 16-poles, LO current 32-pole and the NLO current octupole. The former three multipole moments were previously computed in \cite{buo2} which derived the LO multipoles linear in spin to all orders in the multipole expansion, whereas the NLO current octupole linear in spin is derived here for the first time. See Table \ref{tabl_waveformspin} below for a clear display of which multipole moments are needed for the waveform to 2.5PN. We also reproduce the recently computed \cite{buobla} spin contribution from the tail effect at 2.5PN. 

Together with numerical simulations \cite{numgr}, the results reported here and elsewhere \cite{srad,nrgrs,nrgrproc,eih,comment,nrgrss,nrgrs2,nrgrso,rad1} provide the building blocks for the construction of more accurate GW templates to be used in the data analysis of rapidly rotating compact binaries, e.g. \cite{pan,buo3,hughes}.\\

This paper is organized as follows. In the next section we briefly review the main ingredients for our computation of the spin components in the GW amplitude within the  EFT formalism. In the ensuing section we provide the results for the remaining spin induced source multipole moments, which contribute to the waveform to 2.5PN order. Finally, we discuss nonlinear gravitational effects in the GW amplitude and reproduce the tail effect due to the interaction between the current quadrupole and the mass monopole, which is linear in spin and also enters at 2.5PN order.  A detailed discussion of the formalism to calculate nonlinear effects in the waveform will be elaborated upon in \cite{tail}.

We will rely heavily on previous work and thus encourage the reader to consult the literature on the subject for more details \cite{anditt,srad,nrgr,nrgrs,nrgrproc,eih,comment,nrgrss,nrgrs2,rad1}. Throughout the paper we work in $\hbar=c=1$ units, with Newton's constant and the Planck mass related by $G= 1/( 32\pi m_p^2)$.

\section{Gravitational waveform, multipole moments and spin in NRGR}

\subsection{Gravitational waveform}

The gravitational waveform has a complicated structure due to the nonlinear nature of general relativity.  However the leading effects in the PN expansion are given by the linear propagation of GWs emitted by the source multipole moments.  The waveform in the linearized radiation theory is given by\footnote{The overall minus with respect to the equivalent expression in \cite{thorne,blanchet} is due to the $(+,-,-,-)$ metric signature we use.} \cite{thorne,blanchet} (see \cite{chad, anditt} for its derivation in the EFT framework)
\begin{align}
\label{htt}  h^{TT}_{ij}(t,{\bf x}) & = -\frac{4G}{|{\bf x}|}\Lambda_{ij,kl}\left[\left(\frac{1}{2}\frac{d^2I^{kl}}{dt^2} + \frac{1}{6} \frac{d^3 I^{klm}}{dt^3} n_m + \frac{1}{24}\frac{d^4 I^{klmn}}{dt^4} n_nn_m + \cdots \right) \right. \\ 
& \hspace*{73pt} - \left.    \epsilon^{ab(k} \left(\frac{2}{3} \frac{d^2J^{l)b}}{dt^2} n_a + \frac{1}{4} \frac{d^3J^{l)bm}}{dt^3} n_a n_m +\frac{1}{15} \frac{d^4J^{l)bmn}}{dt^4} n_a n_m n_n  \right. \right. \notag \\
& \hspace*{110pt} {} \left. \left.  +\frac{1}{72}  \frac{d^5 J^{l)bmnp}}{dt^5} n_a n_m n_n n_p + \cdots \right)\right]  \notag 
\end{align}
where the time derivatives of the multipole moments are evaluated at retarded time $t_{\rm ret} = t-|{\bf x}|$, $n_i$ is the unit vector in the direction of observation and  $\Lambda_{ij,kl} = P_{ik}P_{jl} - \frac{1}{2} P_{ij}P_{kl}$ is the transverse traceless projector with $P_{ij} = \delta_{ij}-n_i n_j$ and $(i,j)$ implies symmetrization.

As we discuss momentarily, this expression is sufficient for computing all of the spin effects in the  waveform to 2PN and a sub-set of the 2.5PN contributions.  At 2.5PN however,  nonlinearities in the GW propagation first  contribute to the spin components of the waveform via the tail effect (scattering off the geometry), previously computed in the literature \cite{blandam,blansch,blantail,buobla}. We will come back to this effect towards the end of the paper, and in more detail in \cite{tail}.

\subsection{Multipole moments in terms of the stress energy pseudo-tensor}

The source multipole moments in the EFT formalism were given to all orders in terms of moments of the stress energy pseudo-tensor $T^{\mu \nu}$ in \cite{anditt}, where they were shown to coincide with the long wavelength expansion of the previously derived expressions for the source multipole moments in the more traditional PN formalism \cite{multipolesBDI}.  For our purposes here the relevant results are 
\bea
J^{ijk}_0&=&  \int d^3{\bf x} \left[\epsilon^{iml} T^{0l} {\bf x}^m{\bf x}^j{\bf x}^k\right]_{\rm STF}\label{J0ijk}\\
J_{1}^{ijk} &=&\int d^3{\bf x} \left[\epsilon^{iml}\left( \frac{2}{45}~\dot T^{mn}{\bf x}^n {\bf x}^j {\bf x}^k{\bf x}^l+  \frac{7}{45}~\dot T^{lk}{\bf x}^m{\bf x}^j{\bf x}^2\right)\right]_{\rm STF}\label{J1ijk}\\
I^{ijkl}_0 &=&\int d^3{\bf x} \, T^{00} \left[{\bf x}^i{\bf x}^j{\bf x}^k{\bf x}^l\right]_{\rm STF}\label{Q0ijkl} \\
I^{ijkl}_1 &=& \int d^3 \mathbf x \left(T^{mm} - \frac{4}{5} \dot T^{0m} \mathbf x^m + \frac{13}{110} \ddot T^{00} \mathbf x^2\right) \left[\mathbf x^i \mathbf x^j \mathbf x^k \mathbf x^l\right]_{\rm STF} \label{Q1ijkl}\\
J^{ijkl}_0 &=&  \int d^3{\bf x} \left[\epsilon^{inm} T^{0m} {\bf x}^n {\bf x}^j {\bf x}^k{\bf x}^l\right]_{\rm STF}\label{J0ijkl}\\
J^{ijklm}_0 &=& \int d^3{\bf x} \left[\epsilon^{ipn} T^{0n} {\bf x}^p {\bf x}^j {\bf x}^k{\bf x}^l{\bf x}^m\right]_{\rm STF}\label{J0ijklm} \, ,
\eea
where the STF prescription only acts on the free indices of the multipole moments.\footnote{Note that the form of $J_{1}^{ijk}$ differs from the expression in \cite{anditt}, however they are physically equivalent, which follows from the conservation of $T^{\mu \nu}$.} 
Each moment of the stress energy pseudo-tensor is then computed from Feynman diagrams yielding $T^{\mu\nu}(t, \mathbf k)$, defined as
\beq
\label{partT}
T^{\mu\nu}(t,{\bf k}) = \int d^3{\bf x} T^{\mu\nu}(t,{\bf x}) e^{-i {\bf k}\cdot {\bf x}},
\eeq
and using the Taylor expansion \cite{rad1,srad}
\beq
\label{taylor}
T^{\mu\nu}(t,{\bf k}) = \sum_{n=0}^{\infty} \frac{(-i)^n}{n!} \left(\int d^3{\bf x} T^{\mu\nu}(t,{\bf x}) {\bf x}^{i_1}\cdots {\bf x}^{i_n}\right) {\bf k}_{i_1}\cdots {\bf k}_{i_n}.
\eeq

\subsection{Overview of spin effects in the EFT formalism} 

In this paper we use the EFT formalism for spinning compact objects developed in \cite{nrgrs, nrgrproc,nrgrss,srad}. The spin tensor, $S_L^{ab}$, is introduced in a local frame defined by the tetrad $e^a_\mu$, which obeys $e^a_\mu e^b_\nu \eta_{ab} = g_{\mu\nu}$ and which is expanded in the weak field regime as $e_\mu^a = \delta^a_\mu + \ldots $. The spin vector is defined as usual: $S_L^{ij} = \epsilon^{ijk} S_L^{k}$. The coupling between gravity and spin reads \cite{nrgrs}
\beq
S_{sg} = -\frac{1}{2} \int d\lambda \, \omega_\mu^{ab} u^\mu S_L^{ab}  \label{eq_Ssg} \, ,
\eeq
where $\omega_\mu^{ab}$ are the Ricci coefficients and $u^\mu \equiv \frac{d x^\mu}{d\lambda}$. In what follows we remove the subscript $L$ for convenience.

\begin{figure}[t]
   \centering
   \includegraphics[width=8cm]{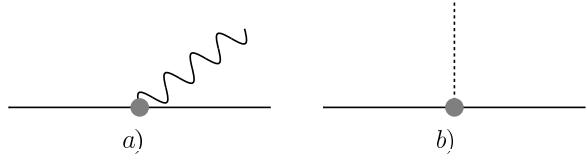}
\caption{\sl ${\cal O}({\bf S}_A)$ worldline coupling to: a) radiation and b) potential modes. A blob represents couplings linear in spin.}\label{world}
\end{figure}

For the radiation sector we follow our previous work \cite{srad}, which we encourage the reader to consult for further details. The calculation of the spin contributions to the source multipoles is divided into two parts: worldline radiation and nonlinear contributions. The Feynman diagram for the radiation stemming solely from one worldline is depicted in Fig. \ref{world}$a$. (Fig.~\ref{world}$b$ represents the coupling of a potential mode to a worldline \cite{nrgr,nrgrs}  which only plays a role as part of a sub-diagram.)  Expanding Eq. (\ref{eq_Ssg}), the interaction Lagrangian between spin and gravity to the order needed for our purposes is~\cite{nrgrs,eih,nrgrss} 
\beq
\label{Lsg}
L_{\rm sg} = \frac{1}{2m_p} h_{\mu\nu,\alpha} v^\mu S^{\nu\alpha} + \frac{1}{4m^2_p}S^{\alpha\beta} h^{\lambda}_\beta \left(\frac{1}{2} h_{\alpha\lambda,\mu} + h_{\mu\lambda,\alpha} -h_{\mu\alpha,\lambda}\right)v^\mu + O(h^3), 
\eeq
where we choose the worldline parameterization $\lambda=t$, such that $u^\mu = v^\mu \equiv \frac{d x^\mu}{dt}$. This yields the Feynman rules \cite{nrgrs,eih,srad,nrgrss} needed for the spin couplings. 

The relevant expressions for the stress energy pseudo-tensor $T^{\mu\nu}(t, \mathbf k)$ are~\cite{srad} 
\begin{align}
 T_{\ref{world}a}^{00}(t, \mathbf k) & = \sum_A S_A^{0i} \, i \mathbf k^i e^{-i {\mathbf k} \cdot \mathbf x_A} \label{eq:WLrad00SA}\\
 T_{\ref{world}a}^{0i}(t, \mathbf k) & = \frac{1}{2} \sum_A \left(S_A^{ij} \, i \mathbf k^j + S_A^{0j} \mathbf v_A^i \, i \mathbf k^j + S_A^{0i} \, i \mathbf k \cdot \mathbf v_A - \dot S^{0i} \right) e^{-i {\mathbf k} \cdot \mathbf x_A} \label{eq:WLrad0iSA}\\
 T_{\ref{world}a}^{ij}(t, \mathbf k) & = \frac{1}{2}\sum_A \left\{\left(S_A^{il} \mathbf v_A^j + S_A^{jl} \mathbf v_A^i \right) i \mathbf k^l + i \mathbf k \cdot \mathbf v_A \left(S_A^{0i} \mathbf v_A^j + S_A^{0j} \mathbf v_A^i \right) \right. \notag \\
& \left. {} \hspace*{39pt} - \dot S_A^{0i} \mathbf v_A^j - \dot S_A^{0j} \mathbf v_A^i - S_A^{0i} \mathbf a_A^j - S_A^{0j} \mathbf a_A^i \right\} e^{-i {\mathbf k} \cdot \mathbf x_A} \label{wlrad},
\end{align} 
where $S^{\mu\nu}$ is the spin tensor that obeys the spin supplementarity condition (SSC) $S^{i0} = \kappa S^{ij}{\bf v}^j+\ldots$ \cite{eih,comment,nrgrs,nrgrs2,nrgrss}, with $\kappa=1$ in the covariant SSC we adopt in this paper. Notice we suppressed the implicit time dependence of the worldline variables. Using Eq. (\ref{taylor}), and expanding in powers of ${\bf k}$ to a given order, we read off the moments of the stress energy pseudo-tensor which enter in the multipoles in Eqs. (\ref{J0ijk}-\ref{J0ijklm}).

Nonlinear effects in the matching for the moments on the other hand are due to radiation off the gravitational part of the stress energy pseudo-tensor which accounts for the potential energy in the binary systems. The relevant diagrams thus involve couplings between radiation and potential modes as shown in Figs. \ref{nonlS9}$abc$. 

\newpage
\subsection{Power counting}

 \begin{table}
  \begin{tabular}{| r | c | c | c |}
    \hline
    {} & ${\cal O}(\not {\hspace*{-2pt}\mathbf S})$ & ${\cal O}({\bf S}_A)$ & ${\cal O}({\bf S}_A^2)$\\ \hline
    $K^{00}_\ell$ & $m r^\ell$ & $m r^\ell v^3$ & $m r^\ell v^4$ \\ \hline
    $K^{0i}_\ell$ & $m r^\ell v$ & $m r^\ell v^2$ & $m r^\ell v^5$ \\ \hline    
    $K^{ij}_\ell$ & $m r^\ell v^2$ & $m r^\ell v^3$ & $m r^\ell v^6$ \\ \hline
  \end{tabular}
  \caption{Scalings for leading terms in moments of $T^{\mu \nu}$ for various orders in spin \cite{srad}. These scalings are valid for all moments $\ell \geq 2$.}
  \label{tabl_Tscalings}
 \end{table}

In order to know at which order spin effects start to contribute to the multipole moments and in turn to the waveform, it is instructive to  derive the scaling of the leading contributions to the moments \beq K^{\mu \nu}_\ell \equiv \int d^3{\bf x}~T^{\mu \nu}(t,{\bf x}) \mathbf x^{i_1} \dots \mathbf x^{i_\ell},\eeq which we summarize in Table \ref{tabl_Tscalings} \cite{srad}. 

The scaling also clarifies which moments of $T^{\mu \nu}$ in the expressions for the multipoles will contribute at a given order. From Table \ref{tabl_Tscalings} and the expressions for the multipole moments\footnote{See \cite{srad, anditt} for the multipole moments beyond the ones in Eqs. (\ref{J0ijk}-\ref{J0ijklm}) needed here.} we see that in order to compute the spin contributions to the waveform to 2.5PN we need, in addition to the results computed in \cite{srad}, only terms linear in spin. It is easy to show that $\calo({\bf S}_A{\bf S}_B)$ terms do not enter until 3PN, and $\calo({\bf S}^2_A)$ contributions are only needed to LO in the mass quadrupole, the mass octupole and the current quadrupole moments \cite{nrgrs,srad}.

In \cite{srad} we computed up to NLO mass and current quadrupole and LO mass and current octupole moments linear in spin. Given the scaling shown in Table~\ref{tabl_Tscalings} we notice that current moments are enhanced by a power of $v$ with respect to mass moments for terms linear in spin. For example, the LO and NLO $I^{ij}$ enter at 1.5PN and 2.5PN in the expansion in Eq. (\ref{htt}), whereas $J^{ij}$ appears already at 1PN and 2PN respectively. 
For completeness we show in Table \ref{tabl_waveformspin} at which orders the spin contributions to the multipole moments enter in the waveform.

 \begin{table} 
  \begin{tabular}{|c | c | c | c |c | c | c |c |} 
    \hline
    order in $h_{ij}^{TT}$ & $J^{ij}$ & $J^{ijk}$ & $J^{ijkl}$ & $J^{ijklm}$ & $I^{ij}$ & $I^{ijk}$ & $I^{ijkl}$\\ \hline
    1PN &  ${\cal O}({\bf S}_A)^{\text{LO}}$ & {} & {} & {} & {} & {} & {}\\ \hline
    1.5PN &  {} & ${\cal O}({\bf S}_A)^{\text{LO}}$ & {} & {} & ${\cal O}({\bf S}_A)^{\text{LO}}$ & {} & {}\\ \hline
    2PN & ${\cal O}({\bf S}_A)^{\text{NLO}}$ & {} & ${\cal O}({\bf S}_A)^{\text{LO}}$ & {} & ${\cal O}({\bf S}_A^2)^{\text{LO}}$ &${\cal O}({\bf S}_A)^{\text{LO}}$ & {}\\ \hline
    2.5PN &  ${\cal O}({\bf S}_A^2)^{\text{LO}}$ & ${\cal O}({\bf S}_A)^{\text{NLO}}$& {} & ${\cal O}({\bf S}_A)^{\text{LO}}$ & ${\cal O}({\bf S}_A)^{\text{NLO}}$ & ${\cal O}({\bf S}_A^2)^{\text{LO}}$ & ${\cal O}({\bf S}_A)^{\text{LO}}$\\ \hline
\end{tabular}
  \caption{Order at which spin contributions to the multipole moments enter the GW amplitude.}
  \label{tabl_waveformspin}
 \end{table}

The remaining multipole moments we compute in this paper are all linear in spin, and consist of the LO $J^{ijkl}$, $J^{ijklm}$ and $I^{ijkl}$ as well as the NLO $J^{ijk}$. Together with the other multipole moments in Table \ref{tabl_waveformspin} already computed in \cite{srad} and the tail contribution involving the leading $J^{ij}$ linear in spin, described in Sec. \ref{tail}, these complete the required ingredients for the GW amplitude to 2.5PN order.

\section{Spin contributions to the source multipole moments}

\subsection{Current octupole $J^{ijk}$}

Given its complexity we begin with the computation of the current octupole. As explained in \cite{srad} we split the Feynman diagrams into worldline and nonlinear contributions.

\subsubsection{Worldline radiation}

We start computing the correction to $J_0^{ijk}$ to NLO from the worldline couplings. Using Eqs. (\ref{J0ijk}) and (\ref{eq:WLrad0iSA}) 
we find
\bea
J^{ijk}_{0,\ref{world}a}&=& \sum_A 2 \left[{\bf S}_A^i {\bf x}_A^j {\bf x}_A^k\right]_{\rm STF}+ \frac{\kappa}{2} \sum_A \left[ 2{\bf v}_A^2 {\bf S}_A^i {\bf x}_A^j {\bf x}_A^k - 2({\bf S}_A\cdot {\bf v}_A) {\bf v}_A^i {\bf x}_A^j{\bf x}_A^k + 4{\bf v}_A^i {\bf v}_A^j{\bf x}_A^k ({\bf S}_A\cdot {\bf x}_A)\right. \nn \\ &-& \left. 4{\bf S}_A^i {\bf v}_A^j{\bf x}_A^k ({\bf x}_A\cdot {\bf v}_A)+({\bf x}_A\cdot {\bf S}_A){\bf a}_A^i{\bf x}_A^j{\bf x}_A^k - ({\bf a}_A\cdot {\bf x}_A) {\bf S}_A^i{\bf x}_A^j{\bf x}_A^k\right]_{\rm STF}\label{j0ijka}.
\eea

We now move to the expression for $J_1^{ijk}$ in Eq. (\ref{J1ijk}). The result reads
\beq
J_{1,\ref{world}a}^{ijk} = -\frac{1}{3}\frac{d}{dt} \sum_A \left[ ({\bf S}_A\cdot {\bf x}_A) {\bf v}_A^i{\bf x}_A^j{\bf x}_A^k- 2 {\bf x}^2_A {\bf x}_A^j {\bf v}^k_A {\bf S}_A^i + ({\bf x}_A\cdot {\bf v}_A){\bf S}_A^i{\bf x}_A^j{\bf x}_A^k- \frac{2}{3}({\bf v}_A\cdot {\bf S}_A) {\bf x}_A^i{\bf x}_A^j{\bf x}_A^k \right]_{\rm STF}\label{j1ijka}.
\eeq

Combining both, Eqs. (\ref{j0ijka}) and (\ref{j1ijka}), we obtain
\bea
J^{ijk}_{0+1,\ref{world}a}&=& \sum_A 2 \left[{\bf S}_A^i {\bf x}_A^j {\bf x}_A^k\right]_{\rm STF}+ \sum_A \left[ \left(\frac{1}{3}-\kappa\right)\left({\bf v}_A^i {\bf x}_A^j{\bf x}_A^k({\bf S}_A\cdot {\bf v}_A)  - {\bf v}_A^2 {\bf S}_A^i {\bf x}_A^j {\bf x}_A^k-2{\bf v}_A^i {\bf v}_A^j{\bf x}_A^k ({\bf S}_A\cdot {\bf x}_A)\right.\right. \nn \\ && \left. +2({\bf x}_A\cdot {\bf v}_A){\bf S}_A^i {\bf v}_A^j{\bf x}_A^k  \right) + \left(\frac{\kappa}{2}-\frac{1}{3}\right){\bf a}_A^i {\bf x}_A^j{\bf x}_A^k ({\bf S}_A\cdot {\bf x}_A)- \left(\frac{\kappa}{2}+\frac{1}{3}\right){\bf S}_A^i {\bf x}_A^j{\bf x}_A^k ({\bf a}_A\cdot {\bf x}_A) \nn \\ && \left.
+ \frac{2}{9} ({\bf a}_A\cdot{\bf S}_A){\bf x}_A^i{\bf x}_A^j{\bf x}_A^k + \frac{2}{3} {\bf x}_A^2 \left( {\bf v}_A^i{\bf v}_A^j{\bf S}_A^k + {\bf x}_A^i{\bf a}_A^j{\bf S}_A^k\right)\right]_{\rm STF},
\eea
for the worldline contribution to the current octupole to NLO.

\subsubsection{Nonlinear gravitational contributions}

\begin{figure}[h!]
    \centering
    \includegraphics[width=8cm]{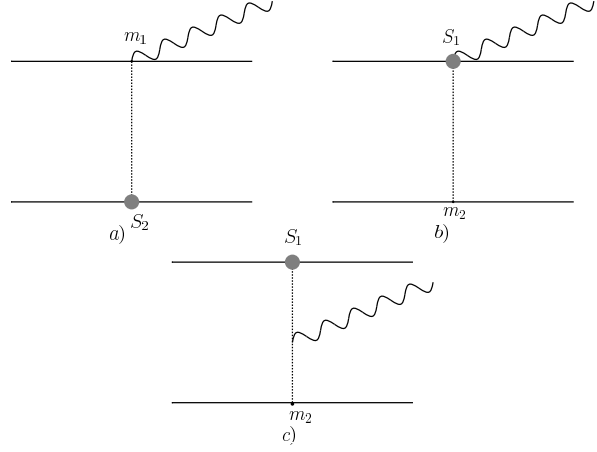}
\caption[1]{First nonlinear diagrams that contribute to $J^{ijk}$ at ${\cal O}({\bf S})$. As explained in the text, the diagram in Fig \ref{nonlS9}a represents a sub-leading effect.}\label{nonlS9}
\end{figure}

The relevant diagrams for the nonlinear gravitational contributions are depicted in Figs.~\ref{nonlS9}$abc$. Since the expression in Eq. (\ref{J0ijk}) entails the $T^{0i}$ component of the stress energy pseudo-tensor, it is easy to show there is no contribution from Fig.~\ref{nonlS9}$a$ at NLO. The same occurred in the computation of the NLO $J^{ij}$ in \cite{srad}. Following the procedure described in \cite{srad} we now compute the remaining diagrams. The contribution from Fig.~\ref{nonlS9}b is relatively simple to calculate and reads
\beq
J^{ijk}_{0,\ref{nonlS9}b}= \sum_{A,B} \frac{G m_B}{2r^3} \left[ ({\bf S}_A\cdot {\bf x}_A) {\bf r}^i {\bf x}_A^j {\bf x}_A^k - ({\bf r}\cdot {\bf x}_A + 8r^2){\bf S}_A^i {\bf x}_A^j {\bf x}_A^k\right]_{\rm STF}\label{j0ijkb}.
\eeq

The most elaborate calculation is that of Fig. \ref{nonlS9}$c$, whose contribution to the stress energy tensor is given by \cite{srad}
\beq
T^{0l}_{\ref{nonlS9}c}({\bf k})=\sum_{A,B} \frac{im_B}{4m_p^2}S_A^{ij}\int \frac{d^3{\bf p}}{(2\pi)^3}(i{\bf p}^j) \frac{-i}{2} \left( {\bf k}^l{\bf k}^i + 2{\bf p}^l{\bf k}^i - \delta^{il} ({\bf k}+{\bf p})^2\right) \frac{-i}{{\bf p}^2}\frac{-i}{({\bf k}+{\bf p})^2} e^{-i({\bf p}+{\bf k})\cdot {\bf x}_B} e^{i{\bf p}\cdot {\bf x}_A}. 
\eeq
In order to use Eq. (\ref{J0ijk}) we need to compute $\int T^{0l}x^mx^jx^k$, which follows from the previous expression by expanding to third order in ${\bf k}$, as shown in Eq. (\ref{taylor}) \cite{rad1,srad}. Performing the integrals, and dropping irrelevant trace terms, we find
\bea
\label{t0l2c}
&&\int d^3{\bf x}~T^{0l}_{\ref{nonlS9}c} {\bf x}^m {\bf x}^j {\bf x}^k = \sum_{A,B} \frac{Gm_B}{r^3} \left[ \frac{1}{3} r^2 S_A^{ml}\left({\bf x}_A^j{\bf x}_A^k+\frac{1}{2}({\bf x}_A^j{\bf x}_B^k+{\bf x}_B^j {\bf x}_A^k)+{\bf x}_B^j{\bf x}_B^k\right) \right.  \\
&-& \left. \frac{1}{3} S_A^{mn} {\bf r}^n {\bf r}^l \left({\bf x}_A^j{\bf x}_A^k+\frac{1}{2}({\bf x}_A^j{\bf x}_B^k+{\bf x}_B^j {\bf x}_A^k)+{\bf x}_B^j{\bf x}_B^k\right) + \left\{m \leftrightarrow j;~m \leftrightarrow k\right\}  - S_A^{ln}{\bf r}^n {\bf x}_B^m {\bf x}_B^k {\bf x}_B^j\right].\nn \eea
Plugging this expression into Eq. (\ref{J0ijk}) we obtain
\bea
J_{0,\ref{nonlS9}c}^{ijk} &=&  \sum_{A,B} \frac{Gm_B}{r^3} \left[\left(r^2 {\bf S}_A^i+\frac{1}{3}({\bf S}_A\cdot {\bf r}){\bf r}^i\right)\left({\bf x}_A^j{\bf x}_A^k+{\bf x}_A^j{\bf x}_B^k+{\bf x}_B^j{\bf x}_B^k\right) \right. \nn\\ && \left.- ({\bf S}_A\cdot {\bf x}_B) {\bf r}^i {\bf x}_B^j {\bf x}_B^k + ({\bf r}\cdot {\bf x}_B) {\bf S}_A^i {\bf x}_B^j {\bf x}_B^k +({\bf r}\times {\bf S}_A)^i ({\bf x}_A\times {\bf x}_B)^j ({\bf x}_A^k+ {\bf x}_B^k)\right]_{\rm STF},  
\eea
which can also be re-written as
\begin{align}
J^{ijk}_{0, \, \ref{nonlS9}c} & = \sum_{A,B} \frac{G m_B}{r^3}\left[\mathbf S_A^i \mathbf x_A^j \mathbf x_A^k \mathbf r \cdot \mathbf x_A + \mathbf r^i \mathbf r^j \mathbf r^k \left(\frac{1}{3}\mathbf S_A \cdot \mathbf r - \mathbf S_A \cdot \mathbf x_B\right) \right. \notag \\
& \left. \hspace*{61pt} + 2 \mathbf r^i \mathbf r^j \mathbf x_{B}^k \left(\mathbf S_A \cdot \mathbf r - \mathbf S_A \cdot \mathbf x_B\right) + \mathbf r^i \mathbf x_{B}^j \mathbf x_{B}^k \left(3 \mathbf S_A \cdot \mathbf r - \mathbf S_A \cdot \mathbf x_B\right)  \right]_{\text{STF}}. 
\end{align}

\subsection{Mass and current 16-pole(s) $I^{ijkl}$, $J^{ijkl}$ \& current 32-pole $J^{ijklm}$}

The remaining multipole moments are needed at LO linear in spin and are given by worldline radiation alone.
The computation follows the same steps as before, and the results are
\bea
I^{ijkl}_{{0+1},\ref{world}a} &=& \sum_A 4\left(\kappa + \frac{3}{5} \right) \left[ ({\bf v}_A\times {\bf S}_A)^i {\bf x}_A^j {\bf x}_A^l {\bf x}_A^k\right]_{\rm STF} - \frac{24}{5} \left[ ({\bf x}_A\times {\bf S}_A)^i {\bf v}_A^j {\bf x}_A^l {\bf x}_A^k\right]_{\rm STF},\\
J^{ijkl}_{0,\ref{world}a}  &=& \frac{5}{2} \sum_A \left[{\bf S}_A^i {\bf x}_A^j {\bf x}_A^k{\bf x}_A^l\right]_{\rm STF},\\ 
J^{ijklm}_{0,\ref{world}a} &=& 3 \sum_A \left[{\bf S}_A^i {\bf x}_A^j {\bf x}_A^k{\bf x}_A^l{\bf x}_A^m\right]_{\rm STF}.
\eea
In the covariant SSC $\kappa=1$, these coincide with the ones given in Eq. (3.6) of \cite{buo2} for the appropriate order $\ell$ in the multipole expansion.

\section{Towards spin effects in the waveforms to 2.5PN order}
Collecting all the ingredients, within the covariant SSC ($\kappa=1$), we finally obtain
\bea
\label{J3f}
J^{ijk}&=& \sum_A 2 \left[{\bf S}_A^i {\bf x}_A^j {\bf x}_A^k\right]_{\rm STF}+ \sum_A \left[ -\frac{2}{3}\left({\bf v}_A^i {\bf x}_A^j{\bf x}_A^k({\bf S}_A\cdot {\bf v}_A)  - {\bf v}_A^2 {\bf S}_A^i {\bf x}_A^j {\bf x}_A^k-2{\bf v}_A^i {\bf v}_A^j{\bf x}_A^k ({\bf S}_A\cdot {\bf x}_A)\right.\right. \\ &+& \left. 2({\bf x}_A\cdot {\bf v}_A){\bf S}_A^i {\bf v}_A^j{\bf x}_A^k  \right) + \frac{1}{6}{\bf a}_A^i {\bf x}_A^j{\bf x}_A^k ({\bf S}_A\cdot {\bf x}_A)- \frac{5}{6}{\bf S}_A^i {\bf x}_A^j{\bf x}_A^k ({\bf a}_A\cdot {\bf x}_A) \nn \\ &+& \left.
\frac{2}{9} ({\bf a}_A\cdot{\bf S}_A){\bf x}_A^i{\bf x}_A^j{\bf x}_A^k + \frac{2}{3} {\bf x}_A^2 \left( {\bf v}_A^i{\bf v}_A^j{\bf S}_A^k + {\bf x}_A^i{\bf a}_A^j{\bf S}_A^k\right)\right]_{\rm STF} \nn \\
&+& \sum_{A,B} \frac{Gm_B}{r^3} \left[\frac{1}{2}({\bf r}\cdot {\bf x}_A-8r^2) {\bf S}_A^i {\bf x}_A^j {\bf x}_A^k + \frac{4}{3} \left({\bf x}_A^i {\bf x}_A^j {\bf x}_A^k-{\bf x}_B^i {\bf x}_B^j {\bf x}_B^k\right){\bf S}_A\cdot {\bf r} - \frac{1}{2} {\bf x}_A^i {\bf x}_A^j {\bf r}^k {\bf S}_A\cdot {\bf x}_A\right]_{\rm STF}\nn,\\
I^{ijkl}&=&\frac{8}{5} \sum_A \left[ 4({\bf v}_A\times {\bf S}_A)^i {\bf x}_A^j {\bf x}_A^k {\bf x}_A^l - 3({\bf x}_A\times {\bf S}_A)^i {\bf v}_A^j {\bf x}_A^k {\bf x}_A^l\right]_{\rm STF},\\
J^{ijkl}&=&  \frac{5}{2}\sum_A \left[{\bf S}_A^i {\bf x}_A^j {\bf x}_A^k{\bf x}_A^l\right]_{\rm STF},\\
J^{ijklm}&=& 3 \sum_A \left[{\bf S}_A^i {\bf x}_A^j {\bf x}_A^k{\bf x}_A^l{\bf x}_A^m\right]_{\rm STF}\label{J5f}.
\eea
These source multipole moments, together with those reported in \cite{srad} and including the necessary spin-independent ones, allow us to obtain the spin dependence in the gravitational waveform to 2.5PN order from Eq. (\ref{htt}) using the equations of motion that derive from \cite{nrgr,nrgrs,eih,comment,nrgrso,nrgrss,nrgrs2}. Still missing is the contribution from  the tail effect in the GW amplitude which we discuss next.

\subsection{Tail effect in the waveform}\label{tail}

Let us now consider the  corrections to the expression in Eq. (\ref{htt}) arising from nonlinearities in the GW propagation. In particular, we are interested in the tail effect which contributes to the waveform linear in spin at 2.5PN. The tail effect for GWs in the PN regime was first studied in \cite{blandam, blansch} where the tail corrections to the waveform were derived. In \cite{rad1} the tail effect was studied in the EFT  formalism, however the results in \cite{rad1} did not include the tail contribution to the waveform. We outline here how to extend the analysis in \cite{rad1} to be applicable to the GW amplitude, and leave the more formal aspects for a future publication \cite{tail}.

\begin{figure}[h!]
    \centering
    \includegraphics[width=5cm]{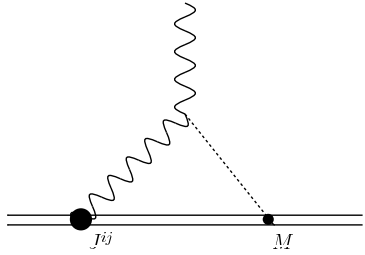}
\caption[1]{Tail diagram that describes the interaction between the source current quadrupole $J^{ij}$ and the monopole $M$. The dashed line here represents the background mode which builds up the geometry sourced by the binary.}\label{tailM}
\end{figure}

 For our purposes we only need to consider the tail produced by the interaction between the current quadrupole $J^{ij}$ and the mass monopole $M$, shown in Fig.~\ref{tailM}. This is because the tail contributes an extra factor of $v^3$ compared to the leading linear contribution from the  current quadrupole $J^{ij}$, cf. Table \ref{tabl_waveformspin}. A similar diagram with $J^{ij} \to I^{ij}$ gives the leading order tail for spinless objects at 1.5PN.\\
 
Our starting point is the expression for the GW amplitude for a given source $T^{ij}$ 
\beq
\label{htt2}
  h_{ij}^{TT}(t,{\bf x}) = -\frac{4 G}{|\bf x|} \Lambda_{ij,kl} \int_{-\infty}^{+\infty} \frac{d\omega}{2\pi} T^{kl}(\omega,\omega {\bf n}) e^{i\omega t_{\rm ret}} + \calo\left(1/|{\bf x}|^2\right).
\eeq
To obtain the waveform we need to compute the corresponding $T^{ij}(\omega,{\bf k}=\omega{\bf n})$, which is defined by means of the GW emission amplitude
\beq 
-\frac{i}{2m_p}  \epsilon^*_{ij}(\omega {\bf n}, h) T^{ij}(\omega,\omega{\bf n}) \equiv i \mathcal A_h(\omega,\omega{\bf n})  = \parbox{21.5mm}{\includegraphics{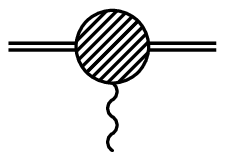}},
\eeq
with $\epsilon_{ij}(\omega{\bf n}, \pm 2)$ the polarization tensor. For the case at hand, we need to compute the diagram in Fig.~\ref{tailM} in order to extract the relevant $T^{ij}$. Moments of the full stress energy pseudo-tensor computed this way, including tail effects, are often called radiative multipoles in the literature \cite{blanchet}.
 
Note that, formally speaking, the expression in Eq.~(\ref{htt2}) is only valid for  sources with compact support. This is not the case for the stress energy pseudo-tensor induced by the $J^{ij}M$ interaction, which scales as 
\beq
 T^{ij} \sim 1/|{\bf y}|^2 + \mathcal O(1/|{\bf y}|^3)
 \eeq
at a distance $|{\bf y}|$ away from the binary's worldline. However, the GW scattering off the long range Newtonian potential induced by the binary at distances far away from the source only leads to a shift in the phase which can be absorbed into an unobservable time redefinition, as will be seen below. For LIGO/Virgo, which are only sensitive to the last few minutes of the binary inspiral, the waveform must be first observed as it enters the frequency band to start tracking its evolution, and it is this change in time which ultimately corresponds to a measurable quantity. This justifies the use of the expansion in Eq.~(\ref{htt2}). See \cite{tail} for a detailed discussion.\\

The tail effect in the graviton emission amplitude for the $I^{ij}M$ interaction was studied in detail in \cite{rad1}, represented by the diagram in Fig.~\ref{tailM} with $J^{ij}\to I^{ij}$. As shown in \cite{rad1} the corresponding integral is IR divergent and can be regularized using dimensional regularization. The result reads
\beq
i {\cal A}_{I^{ij}M}(\omega,\omega{\bf n}) = i {\cal A}_{I^{ij}} \times \left(i G M \omega\right) \left[- {(\omega+ i\epsilon)^2\over \pi \mu^2} e^{\gamma_E}\right]^{(d-4)/2}\times \left[{2\over d-4} - {11\over 6} +\cdots\right],
\eeq
where ${\cal A}_{I^{ij}} = {i\over  4m_p} \omega^2 \epsilon^*_{ij} I^{ij}(\omega)$, and $\gamma_E =0.577\ldots$ is the Euler constant. Note there is a slight difference from the result in \cite{rad1}, i.e. the factor of $(\omega+i\epsilon)^2$ versus $\omega^2+i\epsilon$ in Eq. (18) of \cite{rad1}, because of the necessary retarded boundary condition for the propagator, which takes the form $i/\left((\omega+i\epsilon)^2-{\bf k}^2\right)$ in Fourier space. 

From here we can read off the contribution from the tail to the radiative quadrupole moment by expanding around $d = 4$ (we absorb the factor of $\pi$ into $\mu$ from now on),  
\beq
\label{Itail}
I^{ij}_{\rm rad}(\omega) = I^{ij}_0(\omega) \left\{1+G M \omega\left( {\rm sign}(\omega)\pi + i\left[\frac{2}{\epsilon_{\rm IR}}+\log {\omega^2\over \mu^2}+\gamma_E -{11\over 6}\right]\right)\right\}.
\eeq
Notice the presence of an infinite IR divergent piece proportional to $1/\epsilon_{\rm IR}$ with $\epsilon \equiv d-4$ and the presence of the arbitrary scale $\mu$. 
The IR divergence arises due to the effect of the Newtonian $1/r$ potential sourced by the total (ADM) mass of the binary on the propagation of the GW. 
Similar IR divergences also appear in electrodynamics (QED). As is well known, the $1/r$ Coulomb potential produces an infinite phase shift in scattering amplitudes \cite{weinberg}. However, this is unphysical in the sense that we do not observe the absolute value of the phase shift in QED, and the infinities cancel out of any measurable quantity. See also \cite{weinberg, muzinich} for similar IR effects in gravitational scattering. 
As any IR divergence, the one which occurred here in the computation of the tail cannot contribute to physical quantities. In fact, one can show the IR divergences exponentiate into a pure phase in the emission amplitude, i.e. ${\cal A} = {\cal A}_{\rm IR-finite}~e^{2 i G M \omega/\epsilon_{\rm IR}},$ and therefore drop out of observables extracted from $|{\cal A}|^2$, such as the total power \cite{rad1}. 

 In the waveform on the other hand, the $1/\epsilon_{\rm IR}$ pole remains as an infinite overall phase. However, it is easy to show we can absorb it into a time redefinition, namely
\beq
\label{bare}
 t_{\rm bare} \to \tilde t - 2 G M/\epsilon_{\rm IR},
\eeq
where $t_{\rm bare}$ is the (infinite) bare time, whereas $\tilde t$ represents the time variable used by the experimentalists. This is in accord with our previous discussion, in the sense that one must choose a time
origin to start tracking the signal, which is of course arbitrary. As we will show in detail in \cite{tail}, the IR divergent contribution can always be absorbed into a shift in time.

This renormalization procedure is arbitrary, and that is the reason for the $\mu$ scale in the above expressions. Furthermore, one can easily show that a change $\mu_0 \to \lambda \mu_0$ can be cancelled out by a concurrent transformation $\tilde t(\lambda \mu_0) \to \tilde t(\mu_0) + 2G M \log\lambda$. Hence the GW amplitude is independent of $\mu$.

From the above expressions the contribution from the $I^{ij}M$ coupling to the GW amplitude becomes
\beq
\left(h_{ij}^{TT}\right)_{I^{ij}+I^{ij}M}(\tilde t,{\bf x}) = -\frac{2G}{|{\bf x}|}\Lambda_{ij,kl}\int_{-\infty}^{+\infty}\frac{d\omega}{2\pi}\left[-\omega^2 e^{i\omega \tilde t_{\rm ret}(\mu)} e^{i\theta_{\rm tail}(\omega,\mu)}\left(1+ G M |\omega|\pi\right)\right] I^{ij}_0(\omega),
\eeq
with
\beq
\label{theta1}
\theta_{\rm tail}(\omega,\mu) \equiv G M\omega\left(\log {\omega^2\over \mu^2}+\gamma_E-\frac{11}{6}\right).
\eeq

Up to an unphysical rescaling of $\mu$, this agrees with the result previously obtained in \cite{blansch}. A similar computation gives us the contribution from the current quadrupole and monopole interaction which leads to the leading order spin-dependent effect from Fig.~\ref{tailM}, and reads
\beq
\label{htt3}
\left(h_{ij}^{TT}\right)_{J^{ij}+J^{ij}M}(\tilde t,{\bf x}) = -\frac{2G}{|{\bf x}|}\Lambda_{ij,kl}\int_{-\infty}^{+\infty}\frac{d\omega}{2\pi}\left[-\omega^2 e^{i\omega \tilde t_{\rm ret}(\mu)} e^{i\varphi_{\rm tail}(\omega,\mu)}\left(1+ G M |\omega|\pi\right)\right] J^{ij}_0(\omega),\eeq
with 
\beq
\label{theta2}
\varphi_{\rm tail}(\omega,\mu) \equiv G M\omega\left(\log {\omega^2\over \mu^2}+\gamma_E -{7\over 3}\right).
\eeq
This agrees also with the result in \cite{blantail,buobla}.

Note that while the constants associated with the logarithms in (\ref{theta1}) and  (\ref{theta2}) are unphysical, the difference is not, since we are only free to make one choice for $\mu$. The $\gamma_E$'s are universal (i.e. independent of the multipole) and thus unphysical, but the rational number is non-universal as can be seen from (\ref{theta1}) and (\ref{theta2}), see also \cite{blanchet}. 

The independence of Eq. (\ref{htt3}) from any particular choice for $\mu$ guarantees physical observables do not depend on any specific set of coordinates, as one would expect. However one still has to make a choice, for example $\mu = \omega_s$ (the seismic cutoff frequency for earth-based detectors) \cite{blan2}, which will be directly linked to a reference frequency scale at the time $\tilde t_0$ the GW signal is first observed as it enters the detector's band. Once a choice for $\mu$ is made, and the initial value of the GW phase is measured, the expression in Eq. (\ref{htt3}) tells us how to correlate the predicted evolution of the GW with the observed signal. We will formalize these arguments in greater detail in a future publication \cite{tail}.

\section{Conclusions}

In this paper we computed the  multipole moments, expressed in Eqs. (\ref{J3f} - \ref{J5f}) in the covariant SSC, which together with the results reported in \cite{srad,nrgr, nrgrs,eih,comment,nrgrss,nrgrs2,rad1,nrgrso} lead to the spin contributions in the GW amplitude to 2.5PN order. We employed the EFT formalism developed in \cite{nrgr,nrgrs,nrgrproc,rad1,anditt}, with which we previously computed all the spin-dependent ingredients that enter in the GW phase to 3PN order in \cite{srad}. We also reproduced the leading order spin contribution to the tail effect, already obtained in the literature \cite{blandam,buobla}. In a separate publication \cite{gwamplitude}, we will report the expressions for the GW polarizations with spin effects to 2.5PN order in a ready-to-use form amenable to GW data analysis.
 
\begin{center}
{\bf Acknowledgements}
\end{center} 
We thank an anonymous referee for useful suggestions and for pointing out that \cite{buo2} had previously calculated the leading order multipole moments linear in spin to all orders in the multipole expansion.  This work was supported by: NSF grant AST-0807444, DOE grant DE-FG02-90ER40542, NASA grant NNX10AH14G (RAP) and NASA grant 22645.1.1110173 (AR \& IZR). RAP appreciates the hospitality of the Aspen Center for Physics and Perimeter Institute during the final stages of this work. IZR thanks the Caltech high energy group for its hospitality and the Gordon and Betty Moore Foundation for support.


\begin{thebibliography}{00}

 \bibitem{LIGOVIRGO}
  A.~Abramovici {\it et al.},
  Science {\bf 256}, 325 (1992),
   \url{http://www.ligo.caltech.edu} \\
  A.~Giazotto,
  Nucl.\ Instrum.\ Meth.\  A {\bf 289}, 518 (1990),
  \url{http://www.virgo.infn.it}
  
\bibitem{eLISA} 
  P.~Amaro-Seoane, S.~Aoudia, S.~Babak, P.~Binetruy, E.~Berti, A.~Bohe, C.~Caprini and M.~Colpi {\it et al.},
  arXiv:1201.3621 [astro-ph.CO],
  \url{http://elisa-ngo.org/}

\bibitem{massrange}
C.~Van Den Broeck and A.~S.~Sengupta,
  Class.\ Quant.\ Grav.\  {\bf 24}, 155 (2007)
  [gr-qc/0607092], \
 K.~G.~Arun, B.~R.~Iyer, B.~S.~Sathyaprakash and S.~Sinha,
  Phys.\ Rev.\ D {\bf 75}, 124002 (2007)
  [arXiv:0704.1086 [gr-qc]].
  
\bibitem{parameterestimation}
C.~Van Den Broeck and A.~S.~Sengupta,
  Class.\ Quant.\ Grav.\  {\bf 24}, 1089 (2007)
  [gr-qc/0610126], \
C.~Cutler and M.~Vallisneri,
  Phys.\ Rev.\ D {\bf 76}, 104018 (2007)
  [arXiv:0707.2982 [gr-qc]], \
  E.~K.~Porter and N.~J.~Cornish,
  Phys.\ Rev.\ D {\bf 78}, 064005 (2008)
  [arXiv:0804.0332 [gr-qc]], \
A.~Klein, P.~Jetzer and M.~Sereno,
  Phys.\ Rev.\ D {\bf 80}, 064027 (2009)
  [arXiv:0907.3318 [astro-ph.CO]], \
  C.~Huwyler, A.~Klein and P.~Jetzer,
  arXiv:1108.1826 [gr-qc], \
  P.~Ajith,
  Phys.\ Rev.\ D {\bf 84}, 084037 (2011)
  [arXiv:1107.1267 [gr-qc]].

\bibitem{angularresolution}
  K.~G.~Arun, B.~R.~Iyer, B.~S.~Sathyaprakash, S.~Sinha and C.~V.~D.~Broeck,
  Phys.\ Rev.\ D {\bf 76}, 104016 (2007)
  [Erratum-ibid.\ D {\bf 76}, 129903 (2007)]
  [arXiv:0707.3920 [astro-ph]], \
    M.~Trias and A.~M.~Sintes,
  Phys.\ Rev.\ D {\bf 77}, 024030 (2008)
  [arXiv:0707.4434 [gr-qc]].

\bibitem{numericscomparison}
  M.~Hannam, S.~Husa, U.~Sperhake, B.~Bruegmann and J.~A.~Gonzalez,
  Phys.\ Rev.\ D {\bf 77}, 044020 (2008)
  [arXiv:0706.1305 [gr-qc]], \
  M.~Boyle, D.~A.~Brown, L.~E.~Kidder, A.~H.~Mroue, H.~P.~Pfeiffer, M.~A.~Scheel, G.~B.~Cook and S.~A.~Teukolsky,
  Phys.\ Rev.\ D {\bf 76}, 124038 (2007)
  [arXiv:0710.0158 [gr-qc]], \
  M.~Campanelli, C.~O.~Lousto, H.~Nakano and Y.~Zlochower,
  Phys.\ Rev.\  D {\bf 79}, 084010 (2009)
  [arXiv:0808.0713 [gr-qc]], \
  M.~Hannam, S.~Husa, F.~Ohme, D.~Muller and B.~Bruegmann,
  Phys.\ Rev.\ D {\bf 82}, 124008 (2010)
  [arXiv:1007.4789 [gr-qc]], \
  A.~Buonanno, L.~E.~Kidder, A.~H.~Mroue, H.~P.~Pfeiffer and A.~Taracchini,
  Phys.\ Rev.\ D {\bf 83}, 104034 (2011)
  [arXiv:1012.1549 [gr-qc]].

\bibitem{spinBHs}
  J.~E.~McClintock, R.~Narayan, L.~Gou, R.~F.~Penna and J.~A.~Steiner,
  arXiv:0911.5408 [astro-ph.HE], \ 
  L.~Gou, J.~E.~McClintock, M.~J.~Reid, J.~A.~Orosz, J.~F.~Steiner, R.~Narayan, J.~Xiang and R.~A.~Remillard {\it et al.},
  Astrophys.\ J.\  {\bf 742}, 85 (2011)
  [arXiv:1106.3690 [astro-ph.HE]].

\bibitem{Blanchet:2008je} 
  L.~Blanchet, G.~Faye, B.~R.~Iyer and S.~Sinha,
  Class.\ Quant.\ Grav.\  {\bf 25}, 165003 (2008)
  [arXiv:0802.1249 [gr-qc]].

\bibitem{ampspin} 
  L.~E.~Kidder,
  Phys.\ Rev.\ D {\bf 52}, 821 (1995)
  [gr-qc/9506022], \
  C.~M.~Will and A.~G.~Wiseman,
  Phys.\ Rev.\ D {\bf 54}, 4813 (1996)
  [gr-qc/9608012], \
  K.~G.~Arun, A.~Buonanno, G.~Faye and E.~Ochsner,
  Phys.\ Rev.\ D {\bf 79}, 104023 (2009)
  [Erratum-ibid.\ D {\bf 84}, 049901 (2011)]
  [arXiv:0810.5336 [gr-qc]].
  
\bibitem{nrgr} W. Goldberger and I. Rothstein, Phys. Rev. D {\bf 73}, 104029 (2006)
[arXiv:hep-th/0409156].
 
 \bibitem{nrgrs} R. A. Porto, Phys. Rev. D {\bf 73}, 104031 (2006) [arXiv:gr-qc/0511061].

\bibitem{nrgrproc} R. A. Porto, ``New results at 3PN via an effective field theory of gravity". Proceedings of the 11th Marcel Grossmann Meeting on Recent Developments in Theoretical and Experimental General Relativity, Gravitation, and Relativistic Field Theories, [arXiv:gr-qc/0701106].
\bibitem{rad1}
  W.~D.~Goldberger and A.~Ross,
  Phys.\ Rev.\  {\bf D81}, 124015 (2010). arXiv:0912.4254 [gr-qc].
 
\bibitem{anditt} A.~Ross, 
[arXiv:1202.4750 [gr-qc]].

  \bibitem{dis1} W. Goldberger and I. Rothstein, Phys. Rev. D {\bf 73}, 104030 (2006)
[arXiv:hep-th/0511133].
  
\bibitem{dis2} R.~A.~Porto,
Phys.\ Rev.\  D {\bf 77}, 064026 (2008) [arXiv:0710.5150 [hep-th]].

 \bibitem{LH} For a review see, W. Goldberger ``Les Houches lectures on effective field theories and gravitational radiation." 
Proceedings of Les Houches summer school - Session 86: Particle Physics and Cosmology: The Fabric of Spacetime,  [arXiv: hep-ph/0701129]. Also, R. A. Porto and R. Sturani, ``Scalar gravity: Post-Newtonian corrections via an effective field theory approach"  Ibidem, [arXiv: gr-qc/0701105].

\bibitem{nrgr1}
  W.~D.~Goldberger and I.~Z.~Rothstein,
  Gen.\ Rel.\ Grav.\  {\bf 38}, 1537 (2006)
  [Int.\ J.\ Mod.\ Phys.\  D {\bf 15}, 2293 (2006)] [arXiv:hep-th/0605238].


\bibitem{eih} R. A. Porto and I.Z. Rothstein, Phys. Rev. Lett. {\bf 97}, 021101 (2006) [arXiv:gr-qc/0604099].

\bibitem{comment} R. A. Porto and I. Z. Rothstein,  [arXiv:0712.2032 [gr-qc]].

\bibitem{nrgrs2}
  R.~A.~Porto and I.~Z.~Rothstein,
  Phys.\ Rev.\  D {\bf 78}, 044013 (2008)
  [Erratum-ibid.\  D {\bf 81}, 029905 (2010)]
  [arXiv:0804.0260 [gr-qc]].
\bibitem{nrgrss}   R.~A.~Porto and I.~Z.~Rothstein,
  Phys.\ Rev.\  D {\bf 78}, 044012 (2008)
  [Erratum-ibid.\  D {\bf 81}, 029904 (2010)]
  [arXiv:0802.0720 [gr-qc]].


\bibitem{andi1}  J.~B.~Gilmore and A.~Ross,
  Phys.\ Rev.\  D {\bf 78}, 124021 (2008) [arXiv:0810.1328 [gr-qc]].

\bibitem{nrgrso}
  R.~A.~Porto,
   Class.\ Quant.\ Grav.\  {\bf 27}, 205001 (2010) arXiv:1005.5730 [gr-qc].

\bibitem{perrodin}  
D.~L.~Perrodin,
  arXiv:1005.0634 [gr-qc].

\bibitem{Levi:2010zu} 
  M.~Levi,
  Phys.\ Rev.\ D {\bf 82}, 104004 (2010)
  [arXiv:1006.4139 [gr-qc]].

\bibitem{foffa} S. Foffa and R.~Sturani,
    Phys.\ Rev.\  {\bf D84}, 044031 (2011) arXiv:1104.1122 [gr-qc].


 \bibitem{buo1}  G. Faye, L. Blanchet and A. Buonanno, Phys. Rev. D {\bf 74} 104033 (2006) [arXiv:gr-qc/0605139].

\bibitem{Schafer3pn} J. Steinhoff, S. Hergt and G. Sch\"afer,  
  Phys.\ Rev.\  D {\bf 77}, 081501 (2008) [arXiv:0712.1716 [gr-qc]].

\bibitem{Schafer3pn2} J.~Steinhoff, S.~Hergt and G.~Sch\"afer,
  Phys.\ Rev.\  D {\bf 78}, 101503 (2008) [arXiv:0809.2200 [gr-qc]].

\bibitem{Hergt} S.~Hergt, J.~Steinhoff and G.~Sch\"afer, 
[arXiv:1110.2094 [gr-qc]].
  
\bibitem{Levi:2011eq} 
  M.~Levi,
  arXiv:1107.4322 [gr-qc].
  
\bibitem{Hartung:2011ea} 
  J.~Hartung and J.~Steinhoff,
  Annalen Phys.\  {\bf 523}, 919 (2011)
  [arXiv:1107.4294 [gr-qc]].


\bibitem{buo2} 
  L.~Blanchet, A.~Buonanno and G.~Faye,
  Phys.\ Rev.\ D {\bf 74}, 104034 (2006)
  [Erratum-ibid.\ D {\bf 75}, 049903 (2007)]
  [Erratum-ibid.\ D {\bf 81}, 089901 (2010)]
  [gr-qc/0605140].

\bibitem{srad}  R~A.~Porto, A.~Ross, I.~Z.~Rothstein,
  JCAP {\bf 1103}, 009 (2011). [arXiv:1007.1312 [gr-qc]].

 \bibitem{powerloss} R. A. Porto, A. Ross and I. Rothstein, in progress.

\bibitem{buobla} L.~Blanchet, A.~Buonanno and G.~Faye,
  Phys.\ Rev.\ D {\bf 84}, 064041 (2011)
  [arXiv:1104.5659 [gr-qc]].
 
\bibitem{numgr} L.~Lehner,  
  Class.\ Quant.\ Grav.\  {\bf 18}, R25-R86 (2001).
  [gr-qc/0106072], \
  F.~Pretorius,
  Phys.\ Rev.\ Lett.\  {\bf 95}, 121101 (2005)
  [arXiv:gr-qc/0507014], \
  M.~Campanelli, C.~O.~Lousto, P.~Marronetti and Y.~Zlochower,
  Phys.\ Rev.\ Lett.\  {\bf 96}, 111101 (2006)
  [arXiv:gr-qc/0511048].

  \bibitem{pan}
  Y.~Pan, A.~Buonanno, Y.~b.~Chen and M.~Vallisneri,
  Phys.\ Rev.\  D {\bf 69}, 104017 (2004)
  [Erratum-ibid.\  D {\bf 74}, 029905 (2006)]
  [arXiv:gr-qc/0310034].
  
  \bibitem{buo3}
  A.~Buonanno, Y.~b.~Chen, Y.~Pan and M.~Vallisneri,
  Phys.\ Rev.\  D {\bf 70}, 104003 (2004)
  [Erratum-ibid.\  D {\bf 74}, 029902 (2006)]
  [arXiv:gr-qc/0405090].

\bibitem{hughes}
  R.~N.~Lang and S.~A.~Hughes,
  Phys.\ Rev.\  D {\bf 74}, 122001 (2006)
  [Erratum-ibid.\  D {\bf 75}, 089902 (2007),  Erratum-ibid. D {\bf 77} 109901 (2008)]
  [arXiv:gr-qc/0608062].

\bibitem{tail} R. A. Porto, A. Ross and I. Rothstein, 
 in progress.

 \bibitem{thorne} K.~S.~Thorne,
  Rev.\ Mod.\ Phys.\  {\bf 52}, 299 (1980).


\bibitem{blanchet}
  L.~Blanchet,
  Living Rev.\ Rel.\  {\bf 9}, 4 (2006) [arXiv:gr-qc/0202016].
  
 \bibitem{chad} C.~R.~Galley and M.~Tiglio,
  Phys.\ Rev.\  D {\bf 79}, 124027 (2009) [arXiv:0903.1122 [gr-qc]].
 
 \bibitem{multipolesBDI} 
  T.~Damour and B.~R.~Iyer,
  Phys.\ Rev.\ D {\bf 43}, 3259 (1991), \
  L.~Blanchet,
  Class.\ Quant.\ Grav.\  {\bf 15}, 1971 (1998)
  [gr-qc/9801101].
 
\bibitem{blandam} L.~Blanchet and T.~Damour,
  Phys.\ Rev.\ D {\bf 46}, 4304 (1992).

\bibitem{blansch} L.~Blanchet and G.~Sch\"afer,
  Class.\ Quant.\ Grav.\  {\bf 10}, 2699 (1993).
  
\bibitem{blantail} L.~Blanchet, G.~Faye, B.~R.~Iyer and S.~Sinha,
  Class.\ Quant.\ Grav.\  {\bf 25}, 165003 (2008)
  [arXiv:0802.1249 [gr-qc]].

\bibitem{weinberg}  S.~Weinberg,
  Phys.\ Rev.\  {\bf 140}, B516 (1965).
  
\bibitem{muzinich} I.~J.~Muzinich and M.~Soldate,
  Phys.\ Rev.\ D {\bf 37}, 359 (1988).

\bibitem{blan2}  L.~Blanchet and B.~S.~Sathyaprakash,
  Phys.\ Rev.\ Lett.\  {\bf 74}, 1067 (1995).




\bibitem{gwamplitude} R. A. Porto, A. Ross and I. Z. Rothstein, in progress.

\end{thebibliography}
\end{document}